%% file: ICTIR2023.tex
\documentclass[sigconf,natbib=true]{acmart}
\usepackage{amsmath,nccmath,tabularx} 
\usepackage{algorithm}
\usepackage{algorithmic}
\usepackage{color}
\usepackage{graphicx}
\graphicspath{{images/}}
\usepackage{wrapfig,lipsum}
\usepackage{tabularx}
\usepackage{graphicx}
\usepackage{lscape}
\usepackage{subcaption}
\usepackage{latexsym}
\usepackage{pifont}
\usepackage{multirow}
\newcommand\todo[1]{\textcolor{red}{#1}}

\usepackage{enumitem}
\setlist{leftmargin=3mm}
\usepackage[T1]{fontenc}
\usepackage[utf8]{inputenc}
\usepackage[font=small,labelfont=bf]{caption}

\usepackage{booktabs}
\newcommand{\myparagraph}[1]{\paragraph*{\hspace*{-\parindent}\normalsize\bf#1}}

\usepackage{xcolor}
\usepackage{xspace}
\usepackage{bm}
\usepackage{amsmath}
\usepackage{placeins}

\AtBeginDocument{%
  \providecommand\BibTeX{{%
    \normalfont B\kern-0.5em{\scshape i\kern-0.25em b}\kern-0.8em\TeX}}}

\copyrightyear{2023} 
\acmYear{2023} 
\setcopyright{licensedothergov}\acmConference[ICTIR '23]{Proceedings of the 2023 ACM SIGIR International Conference on the Theory of Information Retrieval}{July 23, 2023}{Taipei, Taiwan}
\acmBooktitle{Proceedings of the 2023 ACM SIGIR International Conference on the Theory of Information Retrieval (ICTIR '23), July 23, 2023, Taipei, Taiwan}
\acmPrice{15.00}
\acmDOI{10.1145/3578337.3605129}
\acmISBN{979-8-4007-0073-6/23/07}

\begin{document}
\title{Exploring the Representation Power of SPLADE Models}


\author{Joel Mackenzie}
\affiliation{%
	\institution{The University of Queensland}
	\city{Brisbane}
	\country{Australia}}
\email{joel.mackenzie@uq.edu.au}
\authornote{Equal contribution}
\author{Shengyao Zhuang}
\affiliation{%
	\institution{The University of Queensland}
	\city{Brisbane}
	\country{Australia}}
\email{s.zhuang@uq.edu.au}
\authornotemark[1]
\author{Guido Zuccon}
\affiliation{%
	\institution{The University of Queensland}
	\city{Brisbane}
	\country{Australia}}
\email{g.zuccon@uq.edu.au}


\begin{abstract}
The SPLADE (SParse Lexical AnD Expansion) model is a highly effective approach to learned sparse retrieval, where documents are represented by term impact scores derived from large language models.
During training, SPLADE applies regularization to ensure postings lists are kept sparse --- with the aim of mimicking the properties of natural term distributions --- allowing efficient and effective lexical matching and ranking.
However, we hypothesize that SPLADE may encode additional signals into common postings lists to further improve effectiveness.
To explore this idea, we perform a number of empirical analyses where we re-train SPLADE with different, controlled vocabularies and measure how effective it is at ranking passages.
Our findings suggest that SPLADE can effectively encode useful ranking signals in documents even when the vocabulary is constrained to terms that are not traditionally useful for ranking, such as stopwords or even random words.
\end{abstract}


\maketitle

\input{sections/introduction}

\input{sections/related_work}
\input{sections/experiments}

\input{sections/results}

\input{sections/conclusion}


\balance
\bibliographystyle{ACM-Reference-Format}
\bibliography{ICTIR2023}

\appendix

\end{document}

%% file: sections/introduction.tex
\section{Introduction and Preliminaries}

Learned sparse retrieval is a paradigm that exploits decades of work on efficient inverted index-based representations {\cite{tmo18-fntir}} and modern deep-learning based relevance modelling by replacing simple statistical payloads (usually term frequencies) with {\emph{learned impacts}} {\cite{l22forum}}.
A number of learned sparse retrieval models have been proposed {\cite{tildev2, arxiv21lm, sigir22mpnl, sigir21mkts, Formal-etal-2021-splade, Formal-2021-spladev2}}; their key differences lie in terms of how or whether document expansion is applied {\cite{tildev1, sigir22mpnl, doct5query}}, the way the model is trained to assign impacts, and the use of query term expansion and weighting.
The SPLADE (SParse Lexical AnD Expansion) family currently represents the most effective approach to learned sparse retrieval, with variations providing different efficiency and effectiveness trade-offs {\cite{Formal-etal-2021-splade, Formal-2021-spladev2, formal2021splade++,lassance2022efficiency}}. 
\citet{nmy23-ecir} provide a unified overview and comparison of various learned sparse retrieval methods.

\begin{figure}[t]
	\centering
	\includegraphics[width=0.9\columnwidth]{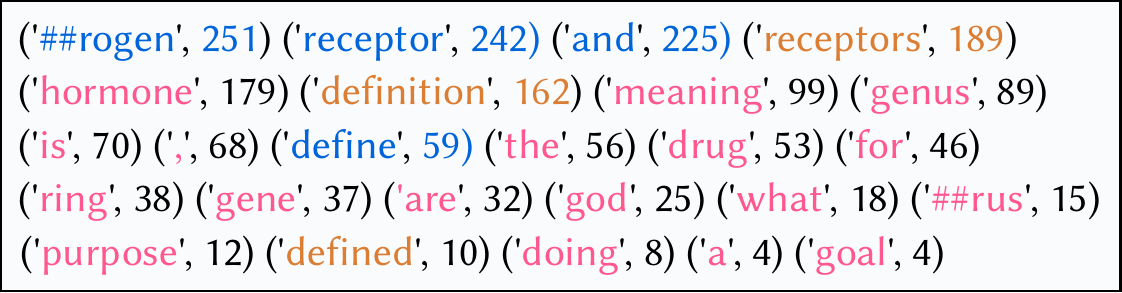}
	\caption{An example of the SPLADE encoding for the query
		{\sf{androgen receptor define}}. 
		Each query term is represented as a {\tt{(`token', weight)}} tuple,
		with blue terms representing the original input query terms, orange
		terms representing alternate inflections on those original terms, and
		pink terms representing entirely new terms. 
		Figure reproduced with permission
		from {\citet{mtl22-tois}}.
		\label{fig-example}}
\end{figure}

\begin{table}[t]
	\caption{Ten most commonly expanded tokens by SPLADEv2 on the 6{,}980 
		MS MARCO {\tt{dev}} queries. These commonly expanded tokens have very
		long postings lists, and are typically not informative.
		\label{tbl-top}}
	\begin{tabular}{r r r r}
		\toprule
		Term & Count & \multicolumn{1}{c}{\%} & List length \\
		\midrule
		{\sf{him}} & 2419 & 34.7 & 2,907{,}631 \\
		{\sf{ring}} & 1995 & 28.6 & 2{,}665{,}265 \\
		{\sf{it}} & 1941 & 27.8 & 2{,}737{,}546 \\
		{\sf{for}} & 1921 & 27.5 & 3{,}116{,}632 \\
		{\sf{a}} & 1838 & 26.3 & 2{,}873{,}128 \\
		{\sf{are}} & 1732 & 24.8 & 2{,}782{,}156 \\
		{\sf{god}} & 1704 & 24.4 & 1{,}293{,}359 \\
		{\sf{.}} & 1659 &  23.8 & 3{,}381{,}977 \\
		{\sf{cause}} & 1657 & 23.7 & 1{,}869{,}128\\
		{\sf{,}} & 1606 & 23.0 & 2{,}553{,}418\\
		\bottomrule
	\end{tabular}
\end{table}

Recently, {\citet{mtl22-tois}} explored the efficiency properties of various learned sparse retrieval models, showing that they are often less efficient than traditional rankers due to so-called ``wacky weight'' distributions.
However, it was not only the distribution of the impact weights that was abnormal; SPLADE was shown to expand terms that seem to have no relevance to the query, including the use of stopwords and punctuation.
Figure~\ref{fig-example} provides an example of this behavior, and Table~\ref{tbl-top} reports the top ten most commonly expanded tokens across the MS MARCO {\tt{dev}} query set.
While it is difficult to understand the reasoning behind these somewhat strange term expansions, we hypothesize that SPLADE may be using tokens that are typically not informative to encode additional signals that can be used during ranking.
As such, the goal of this preliminary study is to better understand why SPLADE expands such terms, and to determine whether SPLADE has the ability to encode signals into typically uninformative lists to improve ranking.

To gain insight into how ``meaningless'' tokens affect the effectiveness of the SPLADE model, we made changes to its MLM head. This allowed the model to expand and assign weights to specific sets of tokens during both training and inference. For instance, we experimented with restricting SPLADE's expansion to a limited set of 150 stopwords (e.g., ``is'', ``the'', and ``and''), which are considered to have little semantic meaning in the context of the content. The surprising outcome was that despite this restriction, SPLADE was still able to perform at the same level of effectiveness as BM25 on MS MARCO dev queries. This result suggests that SPLADE has the ability to encode semantic information effectively, regardless of the tokens used to represent documents. To further understand this phenomenon and explore which types of tokens have the greatest impact on SPLADE's encoding, we tested other settings, such as limiting expansion to low-frequency tokens, and even restricting the vocabulary to  ``latent tokens'' that do not represent {\emph{real}} terms. We believe our findings have the potential to broaden the community's understanding of the SPLADE model, and to motivate efforts to improve the explainability of learned sparse retrieval systems.

%% file: sections/related_work.tex
\section{Related work}
\subsection{SPLADE models}
The SPLADE family of models~\cite{Formal-etal-2021-splade,Formal-2021-spladev2,formal2021splade++,lassance2022efficiency} are BERT-based sparse retrieval models. They have been developed to project every term in queries and documents to a vocabulary-sized weight vector, where each dimension represents the weight of a token in the BERT's vocabulary. This innovative approach estimates the weights of these vectors using the logits of masked language models, and then combines them (e.g., through sum or max pooling) to obtain a single representation of the query and document. This representation can be considered an expansion of the query or document, as it includes terms not originally present. To maintain efficiency, a sparsity regularization loss is applied during training to obtain a sparse representation that can be used more effectively with the inverted index. Compared to other learned sparse methods, such as docT5query~\cite{doct5query}, DeepImpact~\cite{sigir21mkts}, uniCOIL~\cite{arxiv21lm} and TILDEv2~\cite{tildev2}, SPLADE models perform the ``expansion then assigning weights" process in an end-to-end manner. Hence, this model offers a more efficient and effective solution to the vocabulary mismatch problem for learned sparse retrieval.

\subsection{SPLADEv2}
This paper delves into the SPLADEv2 model~\cite{Formal-2021-spladev2}. This model can predict term importance over BERT WordPiece vocabulary ($|V_{BERT}| = 30522$) for any given query and passage. 
SPLADEv2 relies on BERT's contextual embeddings ($h_1, h_2, ..., h_n$) for a given sequence of text tokens (query tokens or passage tokens). Using the MLM head of its trained BERT, it maps each embedding into a weight vector comprising of the entire vocabulary size as follows:
\begin{equation}
\vec{w_i} = MLM(h_i), 
\end{equation}

In this equation, $\vec{w_i} \in \mathcal{R}^V$, where each element $w_{ij}$ within this vector corresponds to the importance score of the $j$th token in the BERT vocabulary. To obtain the final importance score of the $j$th token, a log-saturation effect is applied to the max pooling of all $j$th elements in the weight vectors of the sequence of $n$ tokens. Negative scores are masked out by the ReLU function, resulting in the following equation:

\begin{equation}
	w_j =\max_{i\in n}\log(1 + \text{ReLU}(w_{ij})).
\end{equation}

With the importance score prediction scheme in place, SPLADEv2 is capable of creating a single vocabulary size vector $\vec{q}$ or $\vec{d}$ for the given query or document. Each element in these vectors represents the importance of the corresponding token to the original input text. Following this, the model is trained to learn sparse lexical matching between relevant query-document pairs, using contrastive loss and FLOPS losses. For a more detailed explanation of these losses, we refer the reader to the original paper~\cite{Formal-2021-spladev2}. However, it is crucial to note that, after training, any element in $\vec{q}$ and $\vec{d}$ has the potential to be assigned a high importance score, regardless of whether the token occurs in the original input text or not.



%% file: sections/experiments.tex
\input{sections/results-table.tex}

\section{Experiments}
We now describe our experimental setup, baselines, and the controlled vocabulary mechanisms we explore. Our goal is to examine the effects of different expanded vocabularies on SPLADEv2. To demonstrate the impact, we compare the outcomes with BM25, a dense retriever trained under similar conditions, and the original SPLADEv2.

\subsection{Experimental Setup}

We experiment on the {\tt{MS MARCO-v1}} passage corpus consisting of around $8.8$ million passages {\cite{marcov1}}.
We evaluate our experiments with both the MS MARCO {\tt{dev}} set (containing $6{,}980$ queries and around $1.1$ relevant documents per query) as well as the 2019 and 2020 TREC deep learning track passage ranking queries and judgments ($43$ and $54$ queries for {\tt{DL2019}} and {\tt{DL2020}}, respectively, and both with much deeper judgements) {\cite{trecdl19, trecdl20}}.

\subsection{Baseline Systems}

To situate our findings, we report a few typical baselines as follows.

\myparagraph{BM25}
We use the Pyserini~\cite{Lin_etal_SIGIR2021_Pyserini} toolkit to run BM25 with the tuned parameters $k_1 = 0.82$ and $b = 0.68$ as recommended by {\citet{deepct}}.\footnote{\url{https://github.com/castorini/pyserini/blob/master/docs/experiments-msmarco-passage.md}}

\myparagraph{Dense Retriever}
We train a dense retriever with MS MARCO training dataset by following the example provided in Tevatron DR training toolkit~\cite{Gao2022TevatronAE}.\footnote{\url{https://github.com/texttron/tevatron/blob/main/examples/example\_msmarco.md}}
Specifically, we employed CoCondenser~\cite{gao2022unsupervised} as our backbone encoder model. For each query in the MS MARCO training set, we took the top 200 passages retrieved by BM25 and randomly selected 7 hard negative passages, as well as one positive passage from the qrels. With a batch size of 8, we applied in-batch negatives to each training sample in the batch, which resulted in 63 negatives per training sample. To optimize our training, we set the learning rate to 5e-6 and the number of epochs to 3.

\myparagraph{SPLADEv2}
Our final baseline is the ``default'' SPLADEv2 system. To train SPLADEv2, we follow the example in the Tevatron toolkit from the original SPLADE authors.\footnote{\url{https://github.com/texttron/tevatron/tree/main/examples/splade}}
To ensure a fair comparison with our dense retriever baseline, we utilized the same backbone model and training hyperparameters, and note that all systems were trained on the {\emph{title augmented}} passage collection {\cite{Lassance-2023-marco}}.

\subsection{SPLADEv2 with Controlled Vocabularies}
To determine how important the original representation of each document is for the effectiveness of SPLADEv2, we re-train SPLADEv2 using different vocabularies.
During training, we limit the attention to only tokens appearing in the provided vocabulary; as such, every document and query is represented by just a subset of these vocabulary tokens.
We tested the following vocabularies.

\myparagraph{No Stopword Tokens}
Our first model is a simple modification to the SPLADEv2 baseline discussed above. 
We apply the {\sf{NLTK}}\footnote{\url{https://www.nltk.org/}} English stopword list to the BERT vocabulary to ensure stopwords are not seen by the model during training.
We refer to this approach as {\sf{no-stop}}.

\myparagraph{Stopword Tokens}
Our second approach applies {\emph{only}} the {\sf{NLTK}} English stopword list as the full vocabulary, resulting in a total of $|V| = 150$ unique tokens that can be used to represent each document and/or query.
These tokens are typically considered to be uninformative for ranking and retrieval on English corpora due to their commonality. We refer to this setting as  {\sf{stop-150}}.

\myparagraph{Random Tokens}
We also randomly sample tokens from the original BERT vocabulary ($|V_{BERT}| = 30{,}522$ tokens).
We try both $|V| = 150$ ({\sf{random-150}}) and $|V| = 768$ ({\sf{random-768}}) terms corresponding to the size of the {\emph{stopwords-tokens-only}} and typical dense retrieval dimensionality, respectively.

\myparagraph{Low Frequency Tokens}
Another approach is to simply take the {\emph{lowest frequency}} tokens (the rarest tokens).
In particular, we tokenize the original corpus into BERT tokens and then use the least frequent tokens as the vocabulary,
again experimenting with $|V| = 150$ and $|V| = 768$ (called {\sf{lowfreq-150}} and {\sf{lowfreq-768}}, respectively).

\myparagraph{Latent Tokens}
In this context, we introduce 150 and 768 new tokens that we call latent tokens. These tokens do not exist in the original BERT vocabulary and have randomly initialized embeddings. We conduct experiments in four different settings to explore the impact of these latent tokens.
The first setting is called {\emph{added latent tokens}} ({\sf{added-latent-*}}), where we add our 150 or 768 latent tokens to the original BERT vocabulary, resulting in a total of $|V| = |V_{BERT}| + 150$ and $|V| = |V_{BERT}| + 768$ tokens.
The second setting is called {\emph{latent tokens only}} ({\sf{latent-*}}), where we use our sets of latent tokens as the full vocabulary, resulting in $|V| = 150$ and $|V| = 768$.

%% file: sections/results-table.tex
\begin{table*}[t]
\centering
  \caption{
		Overall effectiveness of the models across all collections.
		The best results are highlighted in boldface.
		Superscripts denote significant differences in paired, two-sided Student's t-test with $p \le 0.01$.
  }	
	\resizebox{0.99\textwidth}{!}{
		\begin{tabular}{c l llllll}
			\toprule
      \multirow{2}{*}{\#} & \multirow{2}{*}{Model} & \multicolumn{2}{c}{\tt{MS MARCO dev}} & \multicolumn{2}{c}{\tt{DL2019}} & \multicolumn{2}{c}{\tt{DL2020}}\\
      \cmidrule(lr){3-4}
      \cmidrule(lr){5-6}
      \cmidrule(lr){7-8}
      & & RR@10 & Recall@1000 & NDCG@10 & Recall@1000 & NDCG@10 & Recall@1000 \\
      \midrule
      A &
			BM25 &
			18.7\hphantom{$^{bcdefghijklm}$} &
			85.7\hphantom{$^{bcdefghijklm}$} &
	  		49.7\hphantom{$^{bcdefghijklm}$} &
				74.5$^{efk}$\hphantom{$^{bcdghijlm}$} &
				 48.8\hphantom{$^{bcdefghijklm}$} &
					80.3$^{efhk}$\hphantom{$^{bcdgijlm}$} 
			\\
					
			B &
			DR &
			\textbf{35.9}$^{adefghijkm}$\hphantom{$^{cl}$} &
			\textbf{97.8}$^{acdefghijkm}$\hphantom{$^{l}$} &
				65.8$^{aefghikm}$\hphantom{$^{cdjl}$} &
			  78.2$^{efghikm}$\hphantom{$^{acdjl}$} &
			  	\textbf{66.1}$^{aefghikm}$\hphantom{$^{cdjl}$} &
					82.4$^{efghikm}$\hphantom{$^{acdjl}$} 
			\\
			C &
			SPLADEv2&
			35.5$^{adefghikm}$\hphantom{$^{bjl}$} &
			97.4$^{adefghikm}$\hphantom{$^{bjl}$} &
				66.9$^{aefghikm}$\hphantom{$^{bdjl}$} &
			  81.1$^{efghikm}$\hphantom{$^{abdjl}$} &
			  	62.3$^{aefhk}$\hphantom{$^{bdgijlm}$} &
					81.9$^{efghikm}$\hphantom{$^{abdjl}$} 
      \\[2ex]
			D &
      {\sf{no-stop}} &
			35.7$^{aefghijkm}$\hphantom{$^{bcl}$} &
			97.5$^{aefghikm}$\hphantom{$^{bcjl}$} &
				\textbf{69.0}$^{aefghikm}$\hphantom{$^{bcjl}$} &
				\textbf{82.8}$^{efghikm}$\hphantom{$^{abcjl}$} &
					64.3$^{aefhk}$\hphantom{$^{bcgijlm}$} &
					\textbf{83.8}$^{efghikm}$\hphantom{$^{abcjl}$} 
			\\
			E &
      {\sf{stop-150}} &
			20.8$^{a}$\hphantom{$^{bcdfghijklm}$} &
			84.4\hphantom{$^{abcdfghijklm}$} &
				42.7\hphantom{$^{abcdfghijklm}$} &
				56.6\hphantom{$^{abcdfghijklm}$} &
					43.9\hphantom{$^{abcdfghijklm}$} &
					62.1\hphantom{$^{abcdfghijklm}$} 
			\\
			F &
      {\sf{random-150}} &
			23.4$^{ae}$\hphantom{$^{bcdghijklm}$} &
			87.0$^{e}$\hphantom{$^{abcdghijklm}$} &
				51.6\hphantom{$^{abcdeghijklm}$} &
				60.0\hphantom{$^{abcdeghijklm}$} &
					50.3\hphantom{$^{abcdeghijklm}$} &
					68.8$^{e}$\hphantom{$^{abcdghijklm}$}
			\\
			G &
      {\sf{random-768}} &
			28.4$^{aefhk}$\hphantom{$^{bcdijlm}$} &
			92.9$^{aefhk}$\hphantom{$^{bcdijlm}$} &
				56.0$^{e}$\hphantom{$^{abcdfhijklm}$} &
				65.1$^{ek}$\hphantom{$^{abcdfhijlm}$} &
					56.7$^{eh}$\hphantom{$^{abcdfijklm}$} &
					72.5$^{eh}$\hphantom{$^{abcdfijklm}$} 
			\\
			H &
      {\sf{lowfreq-150}} &
			24.0$^{ae}$\hphantom{$^{bcdfgijklm}$} &
			87.8$^{aek}$\hphantom{$^{bcdfgijlm}$} &
				47.6\hphantom{$^{abcdefgijklm}$} &
				60.7\hphantom{$^{abcdefgijklm}$} &
					47.9\hphantom{$^{abcdefgijklm}$} &
					66.7\hphantom{$^{abcdefgijklm}$}
			\\
			I &
      {\sf{lowfreq-768}} &
			28.7$^{aefhk}$\hphantom{$^{bcdgjlm}$} &
			93.2$^{aefhk}$\hphantom{$^{bcdgjlm}$} &
				57.3$^{eh}$\hphantom{$^{abcdfgjklm}$} &
				69.3$^{efhk}$\hphantom{$^{abcdgjlm}$} &
					60.2$^{efhk}$\hphantom{$^{abcdgjlm}$} &
					73.5$^{efhk}$\hphantom{$^{abcdgjlm}$}
			\\
			J &
      {\sf{added-latent-150}} &
			35.0$^{aefghikm}$\hphantom{$^{bcdl}$} &
			97.2$^{aefghikm}$\hphantom{$^{bcdl}$} &
				68.2$^{aefghikm}$\hphantom{$^{bcdl}$} &
				79.7$^{efghikm}$\hphantom{$^{abcdl}$} &
					63.0$^{aefhk}$\hphantom{$^{bcdgilm}$} &
					82.2$^{efghikm}$\hphantom{$^{abcdl}$}
			\\
			K &
      {\sf{latent-150}} &
			24.6$^{aef}$\hphantom{$^{bcdghijlm}$} &
			86.6$^{e}$\hphantom{$^{abcdfghijlm}$} &
				53.1$^{e}$\hphantom{$^{abcdfghijlm}$} &
				57.4\hphantom{$^{abcdefghijlm}$} &
					50.3\hphantom{$^{abcdefghijlm}$} &
					67.8$^{e}$\hphantom{$^{abcdfghijlm}$} 
			\\
			L &
      {\sf{added-latent-768}} &
			35.3$^{aefghikm}$\hphantom{$^{bcdj}$} &
			97.5$^{aefghijkm}$\hphantom{$^{bcd}$} &
				68.1$^{aefghikm}$\hphantom{$^{bcdj}$} &
				81.8$^{efghikm}$\hphantom{$^{abcdj}$} &
					64.7$^{aefhk}$\hphantom{$^{bcdgijm}$} &
					83.4$^{efghikm}$\hphantom{$^{abcdj}$} 
			\\
			M &
      {\sf{latent-768}} &
			29.1$^{aefhk}$\hphantom{$^{bcdgijl}$} &
			92.5$^{aefhk}$\hphantom{$^{bcdgijl}$} &
				51.3\hphantom{$^{abcdefghijkl}$} &
				66.1$^{efhk}$\hphantom{$^{abcdgijl}$} &
					60.1$^{aefhk}$\hphantom{$^{bcdgijl}$} &
					73.2$^{efhk}$\hphantom{$^{abcdgijl}$} 
			\\
      \bottomrule
    \end{tabular}
    }
    \label{tbl-all}
\end{table*}

%% file: sections/results.tex
\section{Results}
Table~\ref{tbl-all} presents the effectiveness of the baselines and alternative vocabulary approaches across the MS MARCO
{\tt{dev}}, {\tt{DL2019}}, and {\tt{DL2020}} topics.
We break these results down in the following paragraphs, all in reference to Table~\ref{tbl-all}.

\myparagraph{Are stopwords useful?}
Given that SPLADEv2 was shown to expand seemingly ``unhelpful'' tokens in many queries, we hypothesized that SPLADE may be using these tokens to encode additional ranking signals.
We now try to determine whether this is indeed the case, using stopwords as our proxy for such tokens.
First, let us examine the {\sf{stop-only}} SPLADEv2 (row E) to the baseline systems.
Surprisingly, even if we only allow SPLADEv2 to represent the entire index with $150$ stopwords, it manages to outperform the BM25 baseline on the {\tt{dev}} set with RR@10. 
Although this finding does not hold across the other metrics or collections, it is still surprisingly effective.
This motivated us to further explore how SPLADEv2 leverages stopwords for ranking.

Comparing our SPLADEv2 baseline (B) to the same system {\emph{without}} the inclusion of stopwords during training (D), we observe some (but not significant) improvements when the stopwords were {\emph{removed}}, which contradicts our hypothesis that these tokens are being used by the original model in a meaningful way.

To try to better understand this difference, we also experimented with removing the top 100 most commonly expanded terms on the original SPLADEv2 model during {\emph{query time}} (that is, without re-training the baseline SPLADEv2 model).
Similarly, we observed only a very minor drop in effectiveness (from an RR@10 of $35.5$ to $35.2$ on the
{\tt{dev}} queries).
Overall, this suggests that our original hypothesis does not hold.
So how do we explain the surprising effectiveness of the {\sf{stop-150}} system, which represents all documents and queries through just 150 stopwords?

\myparagraph{Does the vocabulary actually matter?}
Next, we try to understand why training SPLADEv2 on just 150 stopword tokens yields such a surprisingly effective model. Our first idea was to consider alternative vocabularies (with $150$ tokens) -- systems F (random tokens), H (low frequency tokens), and K (latent tokens) -- to see if the findings hold.
Here, we observe even better effectiveness than {\sf{stop-150}}, with all of the alternatives outperforming {\sf{stop-150}} (and rivalling BM25) across most metrics and collections. 
This suggests that SPLADEv2 is capable of modelling useful ranking information in {\emph{arbitrary vocabularies}}, irrespective of the terms within them.

\myparagraph{Increasing the vocabulary size}
Next, we re-examine the controlled vocabulary SPLADEv2 systems by increasing the vocabulary size from 150 to 768 tokens.
Interestingly, we see large (and often significant) improvements (from F to G, H to I, and K to M) just by increasing the vocabulary size.
This is intuitive, because larger vocabularies can provide SPLADEv2 with more ``space'' to accurately model and represent term interactions.
Based on the notion that the vocabulary terms themselves do not seem to make a huge difference to the quality of retrieval, but that the size of the vocabulary seems important, we began to hypothesize whether SPLADEv2 can be thought of as a type of {\emph{dense retriever}}, in the sense that the lexical matching only serves to activate some learned representation for relevance. However, comparing our results to the baseline dense retriever (also using 768 dimensions) demonstrates that these systems (G, I, and M) are significantly less effective than the dense retriever, suggesting that training SPLADEv2 with a constrained representation may have weaker semantic encoding abilities. Despite this finding, it would be premature to conclude that SPLADEv2 cannot function as a dense retriever, especially since additional experimentation is necessary to verify or refute this hypothesis.

\myparagraph{Can latent tokens help SPLADEv2?}
Since even arbitrary lexicons were able to represent documents for ranking tasks, we now test whether {\emph{adding}} tokens to the original SPLADEv2 model can improve performance further. Our intuition is that SPLADEv2 may be able to take advantage of the {\emph{latent tokens}} to form a type of hybrid retriever. During training, we allow SPLADEv2 to use the latent tokens thus increasing the original vocabulary by either 150 or 768.
Rows J and L report the outcome. Interestingly, we find that adding latent tokens can increase the effectiveness of the original SPLADEv2 baseline, though not significantly, suggesting that SPLADEv2 may be able to employ these tokens to its advantage. However, further experimentation is required to understand this phenomenon.


%% file: sections/conclusion.tex
\section{Conclusion and future work}
In conclusion, our study demonstrated that SPLADEv2 is a robust model that can effectively represent documents by a small set of meaningless stopword tokens and still deliver impressive effectiveness. Moreover, our experiments with randomly sampled and low frequency sets of tokens revealed that SPLADEv2 can outperform BM25, even when the expanded tokens are completely irrelevant to the original text. This surprising finding suggests that SPLADEv2 can effectively encode semantic information into any tokens it is given, effectively acting like a dense retriever. Our attempt to make SPLADEv2 act like a hybrid model by adding latent tokens to the original vocabulary showed some improvement in effectiveness, albeit not statistically significant.
Our preliminary study sheds light on the capabilities of the SPLADEv2 model and can help the community gain a deeper understanding of its potential applications. For future work, we suggest to investigate the relationship between vocabulary size (and subset) and SPLADE's representation capabilities. Additionally, exploring the potential use of latent tokens could potentially enhance the semantic encoding power of the SPLADE model. It would also be interesting to examine other models to understand if these findings generalize {\cite{whitebox}}. These avenues of inquiry will help the community to fully comprehend the capabilities of such models and maximize its potential for future applications.

We have made our code publicly available at~\url{https://github.com/ielab/understanding-splade}, allowing others to easily reproduce the results presented in this paper.